\documentstyle[epsfig,10pt]{nnfm}
\begin{document}
\pagestyle{empty}

\title{Gas flow in close binary star systems}

\author{Takuya Matsuda\inst{1}, Kazutaka Oka\inst{1}
\and Izumi Hachisu\inst{2}
\and Henri M.J. Boffin\inst{3}}
\institute{Department of Earth and Planetary Sciences, Kobe University, 
Kobe 657-8501, Japan, \\
\email{tmatsuda@kobe-u.ac.jp}
\and
Department of Earth Science and Astronomy, College of Arts and Sciences, 
University of Tokyo, Komaba, Meguro-ku, Tokyo 153-8902, Japan
\and
European Southern Observatory, Karl-Schwarzschild-Str. 2, D-85748 
Garching-bei-Muenchen, Germany}

\maketitle

\begin{abstract}
We first present a summary of our numerical work on accretion discs in 
close binary systems.
Our recent studies on numerical simulations of the surface
flow on the mass-losing star in a close binary star is then reviewed.

\end{abstract}

\section{Accretion discs}

An accretion disc around a compact star in a close binary star system is an
ubiquitous and essential object. Accretion discs play, for example, important 
roles in cataclysmic variables, nova and X-ray sources. 
The standard theory to explain the physics in accretion discs is the 
$\alpha$-disc model proposed by Shakura and Sunyaev (1973). 
In this theory, the accretion disc is in some kind of 
turbulent state, in which turbulent viscosity is parameterized by a
phenomenological parameter $\alpha$. 

However, the $\alpha$-disc model is rather crude 
approximation to an accretion disc in a close binary system: tidal effects 
due to the companion star are, for example, not taken into account. 
To better take these effects into account, one has to rely on numerical 
simulations.

\bigskip

\subsection{Numerical simulations of gas flow in a close binary system}

A pioneering numerical study of accretion discs in close binary systems was
started by Prendergast (1960). At that time both computers and computational
fluid dynamics were not well developed, so his work was only a preliminary
one. It should be noted that Prendergast also started a pioneering work on
barred galaxies at that time.

Prendergast \& Taam (1974) made a first reliable calculation of gas flow in a
close binary system using the beam scheme developed by Prendergast. The beam
scheme can be considered as a forerunner of the lattice Boltzmann scheme. In
order to solve for the fluid flow, the scheme uses the Boltzmann equation 
rather than the Euler equation as a basic equation. 
The velocity distribution function has values only at fixed points in velocity 
space. The original scheme had the drawback of too much artificial viscosity.

At that time, Sorensen Matsuda \& Sakurai (1974, 1975) were working on a 
numerical study of gas flow in a close binary system, and they were surprised 
to find the paper by Prendergast and Taam. However, since the size of their 
mass-accreting star was very large, their model corresponded to the maybe less 
interesting Algol-type binaries rather than to cataclysmic variables or X-ray 
stars.

Sorensen et al. (1974, 1976) adopted a much smaller size of the
mass-accreting star to simulate a compact star, although the size was still
much larger than that of a realistic compact star, i.e. a white dwarf, a
neutron star or a black hole. If the numerical size of the compact star is
smaller than the so-called circularization radius, an accretion disc may be 
formed.

Sorensen et al. used the Fluid in Cell Method (FLIC) with first order accuracy,
and computed the flow only in the orbital plane, using a Cartesian grid. 
Figure 1 shows the density distribution and velocity vectors of a Roche-lobe 
over-flow in a semi-detached binary system with a mass ratio of one. 
Gas flows out from a mass-losing star (left) through the L1 point towards a 
mass-accreting compact star (right). 
The L1 stream, similar to an elephant trunk is visible but the
accretion disc is not well resolved.

Lin \& Pringle (1976) investigated a similar problem using the sticky
particle method, which utilizes both particles and cells. Particles entering a
cell are assumed to collide, and velocities of particles after the collision
are calculated assuming the conservation of momentum and angular momentum.
This method may be thought as a forerunner of SPH scheme, which is a particle
scheme frequently used in the astrophysics community, but it would be more
appropriate to consider it as a forerunner of the Direct Simulation Monte Carlo
method (DSMC) developed later by Bird (see Bird, 1994). In DSMC, the number of 
particles in a cell is generally much larger, typically 10-100, and collision 
pairs are selected randomly based on collision probability. The present authors
investigated applications of DSMC to astrophysics.

\begin{figure}
  \begin{center}
  \psfig{figure=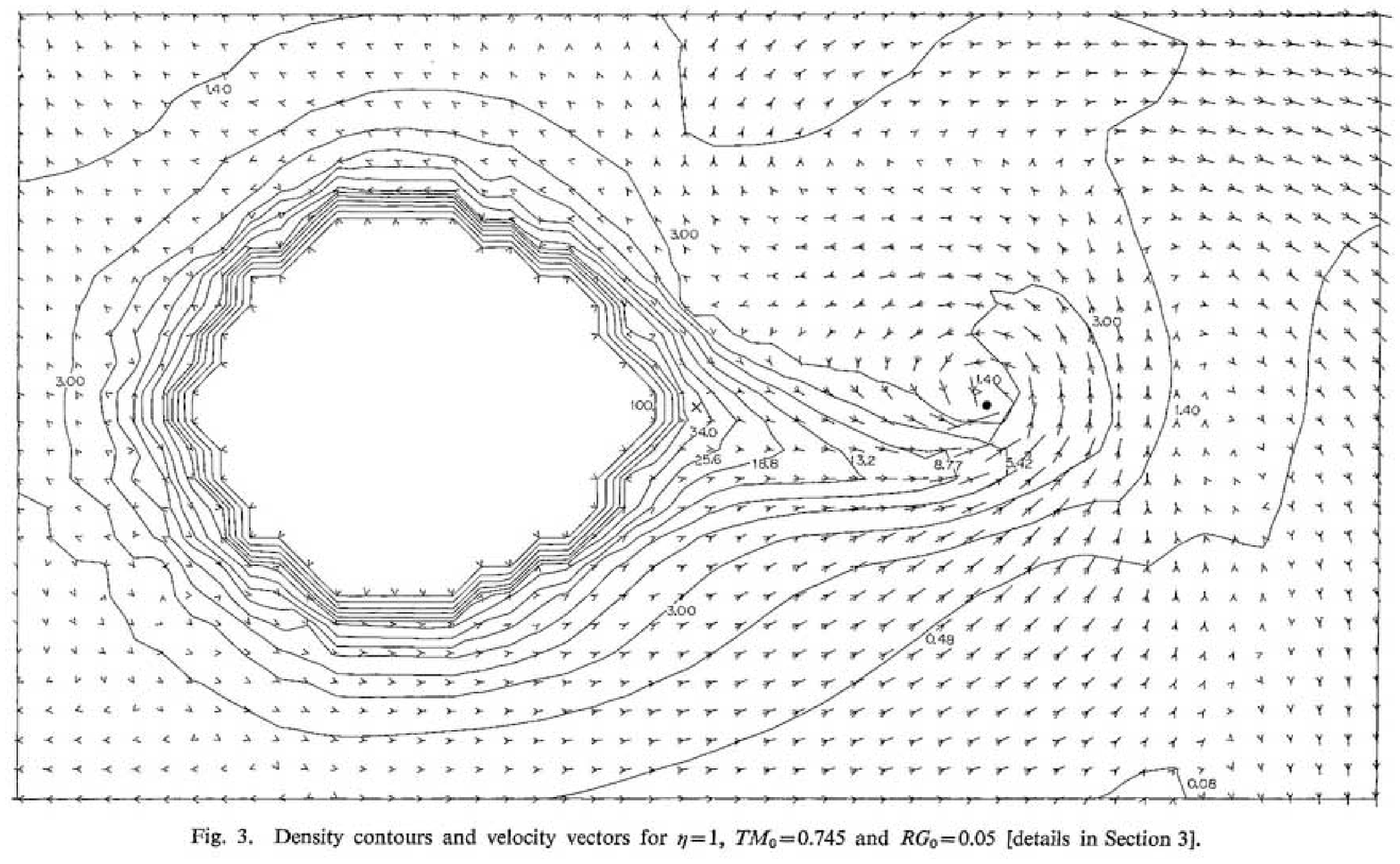,scale=0.35}
  \caption{Two-dimensional hydrodynamic simulation of accretion disc using 
    FLIC method: Density distribution and velocity vectors on the rotational 
    plane are shown. The left oval shape is a mass-losing companion star, 
    while the dot at the right shows the position of a mass-accreting compact 
    object. Gas from the companion star flows through the L1 point towards 
    the compact star due to the gravitational attraction to form a so-called 
    elephant trunk (after Sorensen et al. 1975).}
  \label{Sorensen}
  \end{center}
\end{figure}

\bigskip

\subsection{Modern calculation of accretion flow}

Sawada et al. (1986, 1987) investigated again two-dimensional
calculations of accretion discs using the Osher upwind scheme and Fujitsu
VP200/400 vector supercomputers. Figure 2 shows the density distribution in
the orbital plane in a semi-detached binary system with unit mass ratio. They
first made their calculations using a first-order scheme. When they switched 
to a second-order scheme, they discovered a pair of spiral shaped shock waves, 
as seen in the figure. It is very suggestive that using higher order
scheme reveals a new feature which could not be seen in a scheme with lower 
accuracy.

Spiral shocks in an accretion disc may represent an interesting possibility to
solve a long-standing mystery in the theory of accretion discs, i.e. 
the problem of angular momentum transfer. 
In order for accretion to occur, the gas in the accretion disc has to lose its 
angular momentum. In conventional standard disc model, the disc is supposed to 
be in a turbulent state and the transfer of angular momentum is supposed to 
occur through the turbulent viscosity. However, in spite of many efforts to 
show the disc to be unstable, there has been no success.

In the spiral shock model, gas loses angular momentum at the shocks.
Nevertheless, the spiral shock model had not attracted much attention from 
researchers, and there was even an opinion that spiral shocks did not exist in
three-dimensional calculations. Sawada \& Matsuda (1992) performed the first
three-dimensional hydrodynamic calculation and obtained spiral shocks. Figure
3 shows our recent calculation by Fujiwara et al. (2001). The figure shows an
iso-density surface of an accretion disc around a compact object. Flow-lines on
the iso-density surface and on the orbital plane are visualized by the LIC
method.

\begin{figure}
  \begin{center}
    \psfig{figure=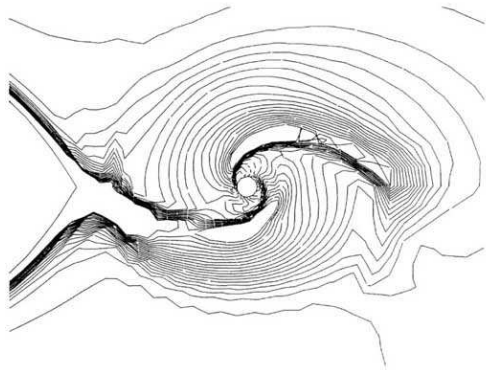,scale=0.5}
    \caption{Calculation based on the second-order Osher scheme: Density 
      distribution on the rotational plane is shown. A circle at the center 
      represents a mass-accreting compact object. Gas from the mass-losing 
      companion (at left) flows through the L1 point and forms an accretion 
      disc. A pair of spiral shock in the accretion disc can be seen 
      (after Sawada et al. 1986, 1987).}
    \label{Figure 2}
  \end{center}
\end{figure}


\begin{figure}
  \begin{center}
    \psfig{figure=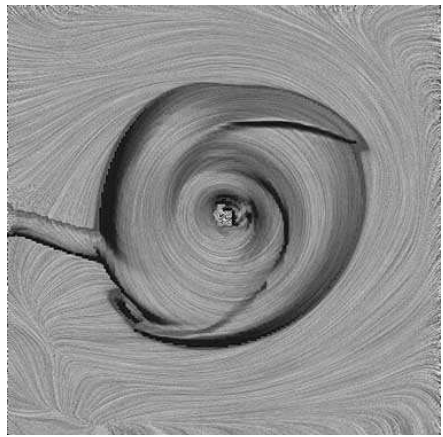,scale=0.3}
    \caption{Recent three-dimensional calculation: Iso-density surface and 
      flow-lines on the surface/rotational-plane are shown. Three-dimensional 
      structure of spiral shocks is evident. It is remarkable that the flow 
      from the L1 point penetrates into the disc (after Matsuda et al. 2000).}
  \end{center}
\end{figure}

\subsection{Discovery of spiral shocks by observation}

As was pointed out earlier, the spiral shock model may solve the long-standing
angular momentum problem. Even if not so, if spiral shocks are present, they
must have some observational implications. In 1997, they were apparently 
detected by Steeghs, Harlaftis \& Horne (1997) in the cataclysmic variable,
IP Pegasi, using the Doppler tomography technique.

Tomography is a technique used, for example, to visualize a cross section of 
the human body by measuring the absorption of irradiated X-rays. In Doppler
tomography, emission lines of hydrogen or helium emitted from hot gas
circulating around a compact star are observed and analyzed to give a Doppler
map. In X-ray tomography, it is the illuminator that rotates about a human, 
but in the case of Doppler tomography, use is made of the rotation of the 
binary system. From temporal variation of the spectrum, one can construct a 
Doppler map, which is a distribution of emission in the velocity space.
From a Doppler map only, it is however not possible to construct uniquely the 
density distribution in the configuration space.
Nevertheless, we may draw useful information from Doppler maps. For
example, if spiral structure of hot region emitting spectrum lines exists, it
is reflected as a spiral structure in the Doppler map. Steeghs et al. found
such a structure. 

The ring-like structure observed in a Doppler map represents an accretion
disc. If the disc is axi-symmetric around a compact star, as is assumed in the
standard disc model, the emission structure should be also axi-symmetric.
However, the emission structure shows a spiral feature.

Interestingly, the surface of the mass-losing companion star is also
bright. This is because the surface of the companion is irradiated by a
radiation from the hot central part of the accretion disc. Moreover this
bright region on the companion start is shifted slightly from a symmetry axis.
It may be due to a current on the surface of the companion star.

\section{Flow on a companion surface}

\subsection{Flow pattern}

So far we discussed the flow in an accretion disc around a compact star, 
because there have been lots of works both theoretical and observational. 
On the other hand the companion star donating gas to the accretion disc has 
not attracted much attention, because it is difficult to observe the flow on 
its surface.  The only exception was the semi-analytic work 
by Lubow and Shu (1975), who predicted the existence of an astrostophic wind 
on the companion surface. But quite recently, surface mapping of the companion 
in cataclysmic variables became possible (see e.g. Dhillon \& Watson 2000 for 
the review).

Oka et al. (2002) performed a three-dimensional simulation of the surface flow 
on the companion and discovered three kinds of eddies associated with a 
high/low pressure on the companion surface: the H-, L1, and L2-eddies. The 
notations H, L1 and L2 denote the high pressure around the pole, the
 low pressure around the L1 point and the low pressure at the opposite side 
to the L1 point, respectively. 

Figure 4a shows streamlines on the surface of the companion star in a 
semi-detached binary system with mass ratio 2; the companion is assumed to be 
two times heavier than the mass-accreting compact star. This mass ratio is 
taken to model a supersoft X-ray source. Figure 4b shows the accretion disc 
for the mass ratio of 2. Note that Figs. 4a and 4b are not the result of one 
calculation. In Fig. 4a the ratio of specific heats, $\gamma$, is assumed to be 
5/3, i.e. an adiabatic gas, while in Fig. 4b, $\gamma=1.01$ is adopted to 
obtain an accretion disc. In order for an accretion disc to be formed, some 
kind of cooling is necessary, and we mimic the cooling by lowering $\gamma$.

\begin{figure}
  \begin{center}
    a)
    \psfig{figure=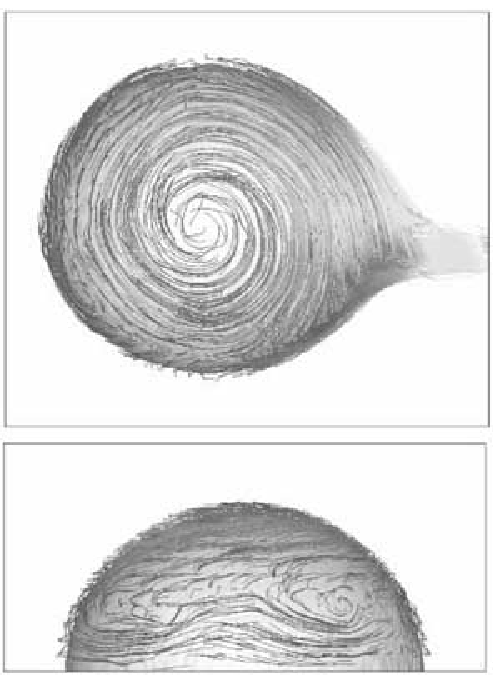,scale=0.8}\hspace{1cm}
    b)
    \psfig{figure=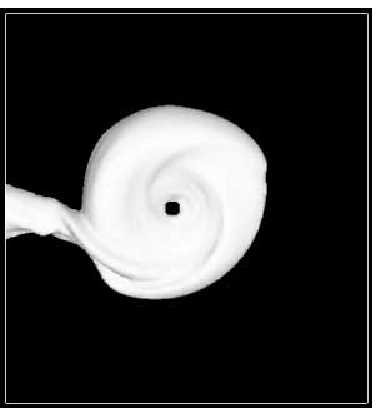,scale=1.3}
    \caption{a) Iso-density surface of the companion star and the streamlines, 
      viewed from the north (top) and those viewed from the negative $x$ 
      direction (bottom). 
      The mass ratio is 2 and the specific heat ratio is $\gamma=5/3$.
      b) Iso-density surface of the accretion disc. The specific heat ratio 
      of $\gamma=1.01$ is adopted.}
    \label{Figure 4}
  \end{center}
\end{figure}


These eddies are nothing but the manifestation of the astrostrophic wind 
predicted by Lubow and Shu (1975). In a rotating fluid, the pressure gradient 
force balances the Coriolis force, and therefore the wind blows along isobaric 
lines. Since gas is withdrawn from the L1 point, a low pressure is inevitably 
formed near the L1 point. Gas near the equator feels less Coriolis force and 
easily flows towards the L1 point, and thus the equatorial region becomes a 
low pressure region. Because of this, a high pressure is formed near the pole 
regions. The mechanism of the formation of the L2 eddy is much more 
complicated.

\subsection{Doppler map}

Based on the above result, we can construct a Doppler map of the surface flow. 
This is not an easy task, because the emission lines are emitted from the 
photosphere of regions of hot temperature. We need to know the temperature 
distribution on the photosphere and have to calculate ionization states of 
either hydrogen or helium on it. Temperature of the photosphere of the 
companion star is very much affected by the irradiation from the central 
region of the accretion disc and the surface of the compact star. Since, 
in the present calculation, we do not take irradiation effect into account, 
we cannot construct real observable Doppler map.

We use the following convention. As a candidate of the photosphere, we
 take an iso-density surface, and plot the horizontal velocity components, 
$V _x$ and $V_y$, of the gas on the $V_x-V_y$ plane, i.e. Doppler map. 

Figure 5 shows the so constructed Doppler map. There are a few characteristics 
to be mentioned. The dark area in and around the companion star is due to the 
gas on the surface of the companion. The fact that it is not restricted within 
the oval shape is reminiscent from the surface flow. If there is no flow, the 
dark region should be within the oval shape. Note that all these dark area is 
not observable, since we do not plot the spectral line intensity in this 
figure. We may argue that the present model Doppler map may be able to explain 
some observational feature in some of supersoft X-ray sources. 
The ring-like structure represents the accretion disc. This shape agrees well 
with observations.

The present model Doppler map is of course a very crude one. We have to 
perform simulations including radiative transfer to construct more realistic 
Doppler map. This is our future task.

\begin{figure}
  \begin{center}
    \psfig{figure=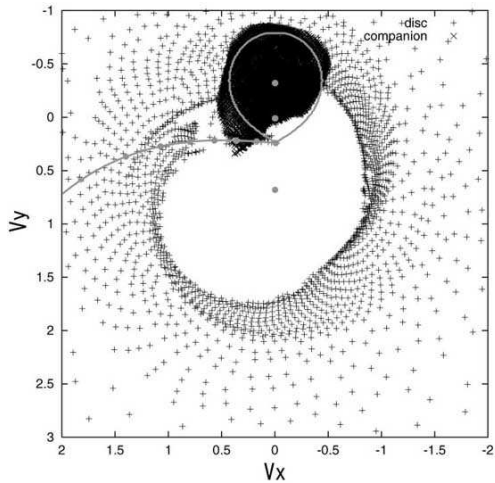,scale=0.7}
    \caption{Constructed Doppler map: $V_x$ and $V_y$ are the horizontal 
      component of the gas velocity. The three dots on $V_x=0$ axis are the 
      center of mass of the companion, the total system, and the compact star, 
      from top to bottom, respectively. The oval shape denotes the companion 
      surface, and the curved line represents a possible ballistic orbit of 
      particles ejected from the L1 point.}
    \label{Figure 5}
  \end{center}
\end{figure}

\section*{Acknowledgements}
T.M. was supported by the grant in aid for scientific research of JSPS 
(13640241). K.O. was supported by the Research Fellowships of the JSPS for 
Young Scientists. This work was supported by "The 21st Century COE Program of 
Origin and Evolution of Planetary Systems" in MEXT.

%


\begin{thebibliography}{5}

\bibitem {Shakura}
N.I. Shakura and R.A. Sunyaev: Astron. Astrophys.. {\bf 24}, 1973, 337.

\bibitem {Prendergas60}
K.H. Prendergast: Astrophys. J. {\bf 132}, 1960, 162.

\bibitem {Taam}
K.H. Prendergast and R.E. Taam: Astrophys. J. {\bf 189}, 1974, 125.

\bibitem {Sorensenn74}
S.-A. Sorensen, T. Matsuda and T. Sakurai: Prog. Theor. Phys. {\bf 52}, 1974,
333.

\bibitem {Sorensenn75}
S.-A. Sorensen, T. Matsuda and T. Sakurai: Astrophys. Space Sci. {\bf 33}, 
1975, 465.

\bibitem{Lin}
D.N.C. Lin and J.E. Pringle: Structure and Evolution of Close Binary Systems, 
Proc. IAU Symp. {\bf 73}, 1976, 237. 

\bibitem{Bird}
G.A. Bird: Molecular Gas Dynamics and the Direct Simulation of Gas Flows, 
Clarendon Press, Oxford, 1994.

\bibitem{Sawada86}
K. Sawada, T. Matsuda and I. Hachisu: Mon. Not. R. Astron. Soc. {\bf 219}, 
1986, 75

\bibitem{Sawada87}
K. Sawada, T. Matsuda , M. Inoue and I. Hachisu: Mon. Not. R. Astron. Soc. 
{\bf 224}, 1987, 307.

\bibitem{Sawada92}
K. Sawada and T. Matsuda: Mon. Not. R. Astron. Soc. {\bf 255}, 1992, s17.

\bibitem{Fujiwara}
H. Fujiwara, M. Makita, T. Nagae and T. Matsuda: Prog. Theor. Phys. {\bf 106}, 
2001, 729

\bibitem{Matsuda}
T. Matsuda, M. Makita, H. Fujiwara, T. Nagae, K. Haraguchi, E. Hayashi and 
H.M.J. Boffin: Astrophys. Space Sci. {\bf 274}, 2000, 259.

\bibitem{Steeghs}
D. Steeghs, T. Harlaftis and K. Horne: Mon. Not. R. Astron. Soc. {\bf 290}, 
1997, L28.

\bibitem{Lubow}
S.H. Lubow and F.H. Shu: Astrophys. J. {\bf 198}, 1975, 383.

\bibitem{Dhill}
V.S. Dhillon and C.S. Watson: Proc. of the Astrotomography Workshop,
Brussels, July 2000, ed. H. Boffin, D. Steeghs (Springer-Verlag Lecture Notes 
in Physics), {\bf p 94}, 2000

\bibitem{Oka}
K. Oka, T. Nagae, T. Matsuda, H. Fujiwara and H.M.J.Boffin: Astron. Astrophys. 
{\bf 394}, 2002, 115.

\end{thebibliography}
\end{document}